\documentclass[acmsmall,anonymous=False]{acmart}
\AtBeginDocument{%
  \providecommand\BibTeX{{%
    \normalfont B\kern-0.5em{\scshape i\kern-0.25em b}\kern-0.8em\TeX}}}

\setcopyright{rightsretained}
\acmJournal{PACMHCI}
\acmYear{2022} \acmVolume{6} \acmNumber{CSCW2} \acmArticle{402} \acmMonth{11} \acmPrice{}\acmDOI{10.1145/3555127}

\begin{document}

\title{Feedback Exchange and Online Affinity: A Case Study of Online Fanfiction Writers}

\author{Ruijia Cheng}
\authornote{Both authors contributed equally to this research.}
\email{rcheng6@uw.edu}
\orcid{0000-0002-2377-9550}
\author{Jenna Frens}
\authornotemark[1]
\email{jfrens@uw.edu}
\affiliation{%
  \institution{University of Washington}
  \city{Seattle}
  \state{Washington}
  \country{USA}
  \postcode{98195}
}

\renewcommand{\shortauthors}{Cheng and Frens}

\begin{abstract}
Feedback is a critical piece of the creative process. Prior CSCW research has invented peer-based and crowd-based systems that exchange feedback between online strangers at scale. However, creators run into socio-psychological challenges when engaging in online critique exchange with people they have never met. In this study, we step back and take a different approach to investigate online feedback by casting our attention to creators in online affinity spaces, where feedback exchange happens naturally as part of the interest-driven participatory culture. We present an interview study with 29 fanfiction writers that investigated how they seek feedback online, and how they identified and built connections with feedback providers. We identify four distinct feedback practices and unpack the social needs in feedback exchange. Our findings surface the importance of affinity and trust in online feedback exchange and illustrate how writers built relationships with feedback providers in public and private online spaces. Inspired by the powerful stories we heard about connection and feedback, we conclude with a series of design considerations for future feedback systems\textemdash namely addressing a range of social needs in feedback, helping feedback seekers signal interests and identity, supporting authentic relationships in feedback exchange, and building inclusive, safe community spaces for feedback.
\end{abstract}




\maketitle

\section{Introduction}

Seeking feedback is a common, important practice in the creative process. Increasingly, creators in many domains use the internet to find feedback that they may not be able to access IRL (in real life) \cite{foong2017online}. HCI and CSCW researchers have studied the mechanism of feedback exchange both in educational \cite{kulkarni2013peer, cook2019guiding} and informal online settings \cite{cheng_critique_2020, kou2017supporting, kotturi2021unique, yen2016social}, inventing scalable systems where non-expert peers or crowdworkers compose feedback for creative work assigned to them \cite{kulkarni2015peerstudio, luther2015structuring, xu2014voyant, cambre2018juxtapeer, ngoon2018interactive}. Despite these efforts, creators run into socio-psychological challenges when engaging in online feedback exchange with people whom they have never worked with IRL \cite{foong2017online}. For example, they may feel uncomfortable and vulnerable in sharing early-stage work to strangers in the first place \cite{kotturi2019designers}. They also face challenges to summarize their context and feedback needs in a short text blurb to someone unfamiliar with their work \cite{cheng_critique_2020}. Additionally, creators are often uncertain about the extent to which they can trust feedback from strangers without knowing their expertise and background \cite{dow2013pilot}. 

We have been reflecting on the current state of online feedback exchange systems. What is missing in systems where feedback is provided by online strangers, even when the exchange is carefully directed? Is there a solution that effectively addresses creators' socio-psychological concerns in online feedback exchange? To further explore these speculative questions, we cast our attention to creators in online \textit{affinity networks}. Online affinity networks, derived from the concept of \textit{affinity spaces} \cite{gee2005semiotic}, are networks of people across platforms interacting over shared interests, identities, and culture. Fueled by shared affinity, diverse networks of people participate in content creation and social interaction, advancing knowledge through socially-situated learning \cite{ito2018affinity}. In contrast with crowd-based and peer-based feedback systems where feedback is routed between users in exchange for currency or class credit, online affinity networks support spontaneous feedback exchange as part of the participatory culture \cite{jenkins2015participatory}. What can we learn from feedback exchange in online affinity networks?

This case study examines feedback exchange in an online affinity network of fanfiction writers. Since the earliest days of the internet \cite{bury2005cyberspaces}, fanfiction communities have used various internet communication channels and sites to create spaces for sharing fanfiction and giving and receiving feedback. Millions of writers have shared fanfiction on archival websites such as Fanfiction.net (FFN) \footnote{\url{https://www.fanfiction.net/}} and Archive of Our Own (AO3) \footnote{\url{https://archiveofourown.org/}}. They also post public reviews to encourage and mentor each other \cite{campbell2016thousands, evans2017more, magnifico2015words}. In addition to reviews on archival sites, prior research has shown, though not explored in depth, that these feedback relationships extend across many other public and private online channels \cite{littleton2011role, aragon2019writers}. We aim to unpack fanfiction writers' feedback practices to understand how online affinity supports feedback exchange and identify design implications for future feedback systems. We thus ask the following research questions:
\newline \newline
\textbf{RQ1:} \textit{How do fanfiction writers seek feedback in online affinity networks?}
\newline
\textbf{RQ2:} \textit{How do fanfiction writers identify and build connections for feedback?}
\newline

%Fanfiction.net alone supported the exchange of 176 million public fanfiction reviews over a decade and a half. Beyond fanfiction-specific sites, their participation crosses several social channels, online spaces of information and social exchange, such as social media, group and private chat, discussion forums, etc. 

%Individuals in fanfiction writing communities have overcome social and technical challenges to seeking feedback by leveraging their affinity networks within an online media ecology. 

To answer these research questions, we interviewed 29 fanfiction writers who have experience participating in online fanfiction communities. We uncovered how writers seek feedback using a variety of online social channels and how they identify and build relationships with feedback providers. This work contributes a deepened understanding of how social needs are met during feedback exchange in different stages in the creative process, how feedback relationships develop and move across spaces, and how the ecology of online platforms supports feedback exchange. We discuss design implications for cultivating connection and relationship-building in feedback systems.

%RQs: 
%%What do fanfic authors look for in feedback and where do they find it
%%How do fanfiction writers seek and build relationships with their feedback providers?

\section{Related Works}
\subsection{Online Feedback Exchange}
%Feedback exchange literature - reviewing existing feedback systems 
%%feedback exchange is a design space, but still has a lot of issues

Feedback exchange undoubtedly plays an important role in the process of producing creative work across fields ranging from interaction design \cite{kou2017supporting} to creative writing \cite{campbell2016thousands}. As a practice first emergent (and first studied) in face-to-face studios and workplaces, creators seek feedback from instructors and peers to improve their work, extend their knowledge and creative skills \cite{butler1995feedback, PUUSTINEN20091014}, and test their work with target audiences \cite{blair2006perception, dannels2008beyond, fasli2017rotational, reily2009two}.

To address the growing need for more, better, and faster feedback, the HCI community has designed innovative peer-based and crowd-based feedback systems. For example, researchers have designed systems for in-person or online classes, scaffolding students to provide feedback to each other with rubrics \cite{kulkarni2015peerstudio}, expert's knowledge \cite{shannon2016peerpresents, ngoon2018interactive}, and structured micro tasks \cite{cambre2018juxtapeer}. Researchers have also developed scalable crowd-based systems that prompt crowdworkers with domain-knowledge support \cite{luther2015structuring}, leverage machine learning methods to control feedback quality \cite{krause2017critique}, and structure feedback with rubrics made by experts \cite{yuan2016almost}.

Despite a growing abundance of research on online feedback systems, socio-psychological dynamics between feedback seekers and providers have not been fully understood \cite{foong2017online}. There are mixed signals regarding the effect of anonymity and criticism on feedback exchange. On the one hand, allowing feedback providers to stay anonymous can encourage specific critical suggestions \cite{hui2015using}; on the other hand, criticism from anonymous feedback providers may impact how creators interpret feedback, as positive affect \cite{nguyen2017fruitful}. There are also challenges of motivating feedback providers to put authentic effort into a stranger's creation \cite{kulkarni2015peerstudio}.

In recent years, because of the expansion of the creative industry and the resulting demand for feedback, exchanging feedback over the internet has become an increasingly common practice for creators, especially those who do not have access to high-quality feedback resources IRL \cite{foong2017online, hui2015using}. Additionally, open questions remain regarding naturalistic online feedback exchange. Creators engaging in informal, interest-driven content creation face challenges with identifying high-quality and stable sources of feedback \cite{xu2012what, crain2017share}. Feedback from online providers can often be underwhelming in quantity, quality, and concreteness \cite{xu2012what}. Novice creators often feel self-conscious about their abilities and hesitate to put their creative work out for critique \cite{crain2017share, kotturi2019designers, marlow2014rookie}. How do creators overcome these challenges? How do they find a community where they are comfortable asking for feedback? How do they solicit feedback that is best suited to their needs? Facing a network of strangers with diverse backgrounds and expertise, how do creators choose whose input to trust \cite{hui2019distributed, yen2016social}? 

With these questions in mind, we conducted this study to unpack how fanfiction writers leverage their online, interest-driven communities for feedback\textemdash how they interact with other community members, what challenges they face, and whether and how they manage to overcome these challenges\textemdash with the goal of learning from fanfiction communities and identifying research and design opportunities for online feedback more broadly. 

%Furthermore, online feedback exchange is a complex, multi-step cycle, and no feedback systems so far are capable of supporting creators' needs over the end-to-end creative process \cite{foong2017online}.

%As a result, instead of widely adopting peer-based or crowd-based feedback systems, creators seek feedback "in the wild" from online interest-driven communities, where they share their work to a wide audience with similar interests \cite{marlow2014rookie, xu2012what}. Creators benefit from getting encouragement and different perspectives from diverse providers in such communities \cite{campbell2016thousands, kou2017supporting}. Emerging scholarly works have explored feedback interactions in online communities with the hope of guiding the design of feedback systems, for example, leveraging a community of interest-driven creators to solicit advice on creation processes \cite{kim2017mosaic, yen2016social}, and learning from creators' strategies to motivate and prompt community members \cite{cheng_critique_2020}. 

\subsection{Online Affinity and Fanfiction Writers}
%Online Affinity - affinity is the glue that holds communities together -- communities where young people learn important skills. What are the mechanisms / relationship between affinity and learning in the fanfiction community (in the instance of feedback exchange)?

The internet enables people with niche media interests to communicate and connect with others who share their passion, accelerating a transformation of media fandom that blurs the lines between consumption and production \cite{jenkins2006convergence}. People creatively expand media, learning how to do so as they engage with \textit{online participatory cultures}---internet cultures where barriers to creative production are low, support for creators is strong, and informal mentoring is commonplace \cite{jenkins2015participatory}. In addition to transforming media itself, this phenomenon has changed the way researchers define media literacy, as well as media literacy learning \cite{scolari2018transmedia}. 

Core to the online participatory culture is \textit{affinity}, which describes the interests, identities and culture that individuals share. The term ``affinity'' was popularized in the study of informal online learning by James Gee \cite{gee2005semiotic}, who coined \textit{affinity space} in reference to places, often online, that draw people of diverse ages and backgrounds to create content, interact with each other, and share knowledge. Later ethnographic investigation by \citet{ito2018affinity} has described \textit{online affinity networks} in reference to interest-driven networks of people who self-affiliate and interact across the open internet. In online affinity networks, the activities of content creation, learning, and social support are often intertwined. For example, online health communities not only afford members to seek support and exchange information, but also allow them to share and practice expressive writing as they journal their experience and emotion \cite{ma2017write}. Similarly, in addition to creating stories, fanfiction authors provide each other with social support through their personal journeys such as coming out \cite{dym2019coming}.

In this study, we focus on a specific case of an online affinity network \textemdash online fanfiction communities. For decades, fanfiction communities have been studied as prototypical examples of online affinity spaces and networks. \textit{Fanfiction} (which we will also refer as ``fanfic'' or ``fic'') is a form of written transformative artwork characterized by the appropriation of elements from fiction, movies and TV, such as plot lines, characters, and worlds. ``Fanfiction'' as a term is differentiated from ``transformative work'' by its emphasis on the fan, a person who is deeply engaged in the source media and its surrounding ``fandom,'' or community and culture \cite{wilson2021fan}. Importantly, fanfiction may also be characterized by its often subversive discourse about emotion and identity \cite{wilson2016role}. Although this expansive definition can be used as a lens to examine premodern literature, \cite{wilson_medieval_fanfiction} our study focuses on contemporary communities tracing back to fandoms surrounding 1900s media such as \textit{Star Trek}, \textit{Sherlock Holmes} and \textit{Doctor Who}. Producing zines and other transformative artworks, many of these social circles were predominantly women in private spaces producing non-commercial fan media \cite{bacon1992enterprising}.

The rapidly-increasing popularization of fanfiction in the late 20th century exemplifies the transformation in the practice of media consumption itself \cite{jenkins2012textual}. Online fanfiction communities today are a living participatory culture \cite{jenkins2015participatory}, forming supportive networks and communities \cite{dym2020social, dym2019coming} and adapting and inventing new technology \cite{fiesler2016archive}. Participation in online fanfiction communities has been studied as an important means of language and literacy learning \cite{black2009online, aragon2019writers, DeLuca2018SharedPS}, co-production of activist media \cite{allred2021be}, identity expression and development \cite{ito2018affinity, dym2019coming}, and digital media literacy development \cite{scolari2018transmedia}. 

%This learning happens outside of school as they build social lives around digital technologies and practices \cite{ito2013hanging}.

Recent research in the CSCW and HCI community has cast attention on the mentoring and feedback exchange relationship in online fanfiction communities. Expertise in fanfiction spaces is distributed; each individual member brings a different set of skills and knowledge, so that mentorship is a fluid exchange with shifting roles of mentor and mentee \cite{black2009online, davis2021mentorship}. Learners may find the support they need distributed among many people, and they actively process and correlate knowledge from artifacts of work, social interactions, and feedback \cite{campbell2016thousands}. The exchange of public reviews has been identified as a salient form of feedback exchange in fanfiction communities. Public reviews exchanged in online fanfiction websites accumulate to a whole of mentorship that helps fulfill writers' needs for affirmation and constructive critiques, leading to improvements in writing \cite{evans2017more, frens2018reviews}. The ability to manage, respond to, and adjust writing based on public feedback is also considered to be transferable to academic writing and other aspects of life \cite{DeLuca2018SharedPS}. 

While the most recent studies on fanfiction feedback equate feedback exchange with public online reviews, public feedback is just the tip of the iceberg in overall collaboration among writers. Exchanging comments on fanfiction and other fan media was commonplace pre-internet, as was collaboration among social circles of fans in their homes, in mini-conventions (i.e., 50 or so people), and in large conventions where people travelled long distances to physically co-locate \cite{bacon1992enterprising}. During the development of the early internet, women fans began creating private channels online where they could socialize and discuss fandom \cite{bury2005cyberspaces}. Modern fandoms exist across an ecosystem of online spaces, public and private. Besides writing reviews on fanfiction websites, writers participate in close-knit online groups to critique each other's works \cite{littleton2011role}. It is common for fanfiction writers to have their stories proofread by someone before posting online (referred to as ``beta reading'') \cite{kelley2016chocolate}. In addition, writers often engage in collaborative creation of fiction together with other forms of media content \cite{hellekson2009fannish}. These types of behind-the-scene feedback interactions can potentially offer us a new model to study and design feedback exchange. We thus carry out this exploratory study that investigates fanfiction writers' feedback practices in online affinity networks, to better inform our understanding of these inspirational communities and the design of future feedback spaces and systems.

%Prior research has explored how writers communicate identities, build relationships and encourage feedback with their audiences on fanfiction websites by writing to them in author notes \cite{black2009online}. In the present study, we explore why and how authors use a range of social channels involved in feedback-seeking and relationship building.

%Feedback in fanfiction communities varies in its purpose, content and quality \cite{evans2017more, littleton2011role, magnifico2015words}. The \textit{distributed mentoring} framework categorized public fanfiction reviews based on content, with some focused on emotional support, encouragement and socializing, while others focus on constructive suggestions on different levels \cite{evans2017more}. While \citet{littleton2011role}'s work revealed that feedback exchange among fanfiction writers is likely distributed on many corners of the internet, 

%we hope to draw a holistic view of how feedback exchange happens, how fanfiction writers identify and leverage many-to-many, many-to-one and one-to-one channels, and how we can better support feedback exchange knowing that it is distributed on multiple spaces. 

%We believe that there is room to expand prior work on feedback exchange in online ecosystems with respect to fanfiction online informal learning. 

\section{Methods}

\subsection{Participants and Recruitment}
%We conducted 29 semi-structured interviews with fanfiction writers. All participants were active fanfiction writers, with experience writing fanfiction spanning from 18 months to more than 20 years. We recruited participants through a process of purposeful sampling, stratifying across years of fanfiction experience so we would hear from all experience levels, as well as participation in different archives (e.g. Fanfiction.net, Archive of Our Own, WattPad, LiveJournal, Tumblr, FIMFiction.net, and others), based on data from a previous survey that we distributed on Tumblr \footnote{\url{https://www.tumblr.com/}} in November 2017. The survey included demographics and questions on their level of experience in writing fanfiction. Among the 1,888 responses, we sampled recruits with a relatively balanced distribution of writing experience. From November 2018 to February 2019, we sent interview invitation messages to the sampled recruits through emails and Tumblr private messages. In the end, 29 writers responded and participated in the interviews.

We conducted 29 semi-structured interviews with fanfiction writers, sourced from a recruitment survey we distributed on Tumblr \footnote{\url{https://www.tumblr.com/}} in November 2017. In the survey, we asked participants to report demographic information, years of experience in writing fic, and platforms that they used for fanfiction-related activities. Among the 1,888 respondents, we stratified active writers into groups based on their self-reported experience levels, which spanned from 18 months to over 20 years. We then sent interview invitations within each experience level to ensure that we would hear from newer community members as well as those who have been writing fanfiction for decades, which contributes different perspectives of feedback exchange. We also ensured our participant group contained users of a range of different fanfiction platforms (e.g., Fanfiction.net\footnote{\url{https://www.fanfiction.net/}}, Archive of Our Own\footnote{\url{https://archiveofourown.org/}}, WattPad\footnote{\url{https://www.wattpad.com/}}, LiveJournal\footnote{\url{https://www.livejournal.com/}}, FIMFiction.net\footnote{\url{https://www.fimfiction.net/}}, and others), so that our findings could be established on user experience across a diverse set of platforms and communities. We contacted interview participants by email or by Tumblr direct message. As this was unfunded research, we were unable to offer any compensation to participants; this was entirely a volunteer effort.

We collected self-reported demographic information (age, race, nationality, and gender) in the recruitment survey. Since some participants were well known in the community, a combination of their quotes in our study and demographic information might deanonymize them. Therefore, to protect the anonymity of our participants, instead of mapping individual participants to their demographic information in a table, we summarize the profile of our participants as follows. 
Out of the 29 participants, we were able to successfully record demographic information from 26 of them. At the time of the survey (about one year prior to the interviews), 4 participants self-reported their age as 16-18, 4 were age 19-23, 7 were age 24-29, 6 participants were 30-40 years old, 3 participants were 40-50 years old, and 2 participants were 50 or older. In self-reported race, 21 were white, one was Hispanic and white, one was Latinx and white, one was Filipino, one was black/African American, and one declined to provide information about their race. In nationality, 17 were from the United States, 4 were from the United Kingdom, 2 were from Canada, one were from France, one were from Italy, and one were from Australia. In gender, 14 were women and 11 were nonbinary, transgender men, genderqueer, bigender, or agender. %Zero participants self-reported as cisgender men. 
One participant preferred not to disclose their gender.  
%We did not disclose the race of one participant for reasons of anonymity.

%Fanfiction is a prototypical space for the study of online affinity and participatory culture \cite{jenkins2012textual}. Note, however, that people connect over many affinities beyond fandom, and contemporary research describes many domains where people are able to informally create, connect and learn. In the present study, we interviewed fanfiction writers who answered our survey on Tumblr. 
Instead of a single fanfiction archival website or a single fandom, we chose Tumblr, a popular social media platform among fan creators, as a suitable recruitment channel for learning about a cross-section of experiences across fandoms and sites. While we recruited our participants from Tumblr alone, during the interviews we encouraged participants to share experiences about any platforms or channels that they used for feedback. Tumblr is not an all-encompassing fan platform, and fanfiction communities have historically needed to migrate platforms due to policy and other factors. Our interviews occurred in January to February 2019, during a time when migration away from Tumblr was ongoing after the infamous December 2018 NSFW content purge \footnote{\url{https://fanlore.org/wiki/Tumblr_NSFW_Content_Purge}}. As a result, our participants had experienced multiple platforms in the past, or were actively exploring new ones. Therefore, our analysis develops themes from diverse experiences across many fandoms and fanfiction-related communities.

%TODO: insert participant table
%participant ID, years of experience, major sites and channels

\subsection{Procedures}
Each semi-structured interview was 60 minutes long and was conducted over Google Hangouts by one of the co-first-authors, while the other observed. The interviews focused on three broad areas: participants' practices of seeking feedback on their fanfiction, their connections with feedback providers, and their use of technology in the process of getting feedback. In the optional second part of the interview, participants shared and talked through an experience in which they received feedback they considered valuable. 
%XX participants shared a work through URL or screen-share. 
Interviews were audio or video recorded according to participants' consent. % (we already said this) This research was not funded and all effort on the part of both the researchers and participants was volunteer labor.

\subsection{Data analysis}
The authors of this paper manually transcribed all of the approximately 30 hours interview data. Then, they coded the interview data using a grounded theory methodology as outlined by Charmaz \cite{charmaz2006constructing}. There was an iterative open coding phase followed by a closed coding phase. During the open coding phase, the authors built a qualitative coding set over four iterations of open coding. During each open coding iteration, the authors separately performed generative open coding on the same two transcripts, and then discussed the transcripts line-by-line to come to agreement. After each open coding iteration, they thematically categorized codes, building a hierarchy by using an affinity diagram. After agreeing on completion of the open coding phase, the authors divided the remaining transcripts for a closed coding phase. They separately coded the remaining transcripts and met to discuss any concerns and needed modifications to the coding set. The resulting final codebook contained 131 codes categorized into 9 major themes and 22 sub-themes.

Over the course of the interviews and coding phases, the authors wrote 23 memos to synthesize ideas from the interviews and coding set. Each memo described an emergent theme (i.e., combination of codes) from the study. The authors furthermore performed a thematic pass of each transcript, highlighting and commenting on excerpts that related to the major themes of the study. Over the course of interviewing, transcribing, highlighting, coding, and writing memos, the authors transformed the raw qualitative dataset into a series of themes on social feedback exchange practices in online affinity networks.

\subsection{Researcher Positionality}
Our positionalities as researchers place us adjacent to the community of fanfiction writers. Jenna self-identifies as queer, transgender, autistic, and an avid roleplayer, gamer, and consumer of fan works, as well as a mixed-method researcher of online communities. Ruijia is also a mixed-method researcher of online communities, as well as an active creator and consumer of online fan works. Sharing identities with our participants and familiarity with fandom helped us establish common ground and build trust during the interviews.  

\section{Findings}
\subsection{How do fanfiction writers seek feedback in online affinity networks?}
\label{finding:feedback-practice}
%Types of feedback sought in hierarchical affinity spaces, and how these spaces facilitate/ meet those needs
In this section, we discuss the different social needs in feedback exchange at different stages in the creative process, and how a variety of public or private online channels support these needs. We identify four distinct practices of feedback exchange and name them using direct quotes from participants: ``throw ideas at each other,'' ``give my friends a little snippet,'' ``beta reading,'' and ``all I want for Christmas are reviews.'' 

\subsubsection{``Throw Ideas at Each Other''}
\label{findigns:idea}
%%TODO: write a couple of sentences on what's this space/group chats
The earliest point in fanfiction writing where writers seek feedback is while brainstorming story ideas. It is a common practice for writers to describe half-baked ideas to one feedback provider, or a small group, and bounce ideas. For example, P1 described a typical situation where members of a fanfiction-specific group chat showcased and discussed story ideas with each other: 

\begin{quote}
``Somebody says 'I'm really stuck with this idea,' and they're talking about their idea. So then it's a case of getting them to open up about what the idea is, and what they want to achieve with the story, and how they feel about writing it, and then it becomes a discussion sort of about the mechanics of writing it, and how they're going to do it, and getting them into a position where they feel that they do want to write it, and they feel confident about starting or continuing, or the direction that they want to go in.'' (P1)
\end{quote}

As illustrated in the above example, writers talked through ideas with other writers to identify problems before they became deeply involved in the writing. They discussed strategies, planned for the plot, and got affirmation to continue writing. Our participants reported that even in cases where they themselves had no direct need for feedback, it was valuable to get inspiration from others' ideas: ``it can be inspiring to hear people's ideas, even you don't want to use them, because it just helps spur your own creativity.'' (P18)

We found that group chats (e.g., Discord servers) were particularly felicitous spaces for feedback during writers' ideation stage. Since getting feedback in the ideation stage was a dynamic process that often involved back-and-forth discussion and debate, participants shared that they would prefer informal and immediate exchanges with others. Group chats afforded this type of interaction:

\begin{quote}
``When we were talking about the actual content of the fics we were writing, we would just throw ideas at each other, oh this character did that? It would be really cool if this person reacted in that way. Or I think that's kind of out of character, maybe this should happen instead, kind of thing. Because it was an online group chat, it was just like immediate responses.'' (P20)
\end{quote}

Besides getting dynamic and conversational feedback, participants also reported their preference to engage with multiple people in the ideation process so that they could get comprehensive feedback from different angles: 

\begin{quote}
``[On Discord:] that's a good way to get ideas, good place to discuss ideas and kind of figure out how you want to make something work in a story. And just exploring different aspects and discussing that with people.'' (P25)
\end{quote}

Although getting feedback from lots of people early on was beneficial, many participants considered getting feedback on half-baked ideas to be a very private endeavor. Showing ideas for stories that they themselves are not sure about requires trusting and intimate relationships. For that reason, participants turned to group chat services, which afford non-public conversations with a small group of close friends. With the commonly mentioned group chat service Discord, writers leveraged the ``subchannel'' feature to create different layers of privacy for conversation than in a bigger group. With only a small group of close friends, writers were more willing to open up about their ideas and discuss their thoughts:

\begin{quote}
``If it's to my friend, it's before I'm done or before I've even been drafting it… Because my friends, they are part of the writing process. I bounce ideas off of them, I plot with them, and I help them plot their things. They're part of the feedback cycle from the beginning.'' (P29)
\end{quote}

Group chats also supported social activities that were not necessarily limited only to serious feedback exchange. Social activities, such as joking around about ideas, or even roleplaying characters to flush out thoughts, were also an important part of the brainstorming process:

\begin{quote}
``A lot of what brainstorming together sounds like is us joking around about the characters, or joking around about fanfiction tropes that we really enjoy. We'll be like, and what if this dramatic thing happens. And we'll do a quick little joke roleplay with the characters, that's like, we're joking, the dialogue that we're saying is not going to be involved but a lot of times it gets us thinking about the different possibilities.'' (P17)
\end{quote}

Writers blurred the boundary between socialization and feedback in ideation conversations, as shown in the above example. They brainstormed together with their friends, getting feedback in a casual, interactive, and converging manner.  

\subsubsection{``Give My Friends a Little Snippet''}
After deciding on the story idea, the next stage where writers needed feedback was after they jotted down their ideas into short little snippets of a few sentences to a few paragraphs. Similarly to the ideation stage, writers posted in-progress snippets of their writing to small group chats with close friends who shared the same interest in the fandom. For example, P12 talked about sharing a snippet of their work with a group of friends when they were not sure what to write next: ``Sometimes I will give my friends a little snippet and be like, I'm stuck here what do you think I should do.'' In another example, P9 posted an initial outline of their story to a group of friends and asked for their opinion on the plot line: 

 \begin{quote}
``A lot of times I have issues just flushing out the plot, sort of my biggest issue, so I will just show people my outline and stuff, and I'll talk it through with them and they will help me to flesh out my plot a little bit more.'' (P9)
 \end{quote}
 
When writers felt unsure about some aspects of their writing, they preferred to quickly troubleshoot or get reactions on the in-progress work, rather than to wait until they completed the entire draft. Additionally, sending in-progress work to a group of fandom friends increased the chance of getting quick help and different perspectives: 

\begin{quote}
``I'll post a snippet of a scene and be `hey, how do you guys think about this part? I am working on it right now'. With the discord group they are very immediate. They're really good for in-the-moment help.'' (P15)
\end{quote}

\begin{quote}
``If it's a minor problem, I might just sort of say, `can someone look at this paragraph and tell me what's wrong with it?' So you might throw just that paragraph into the group chat and then everyone can kind of chip in and see what they think.'' (P1)
\end{quote}

Having close relationships with feedback providers eased feelings of vulnerability while sharing and asking for help with in-progress work. Participants described instances where being in a small private group chat dedicated to feedback exchange gave them a sense of safety and closeness, lowering the social anxiety associated with revealing their confusion and weakness. They felt comfortable sharing their uncompleted and unedited work without worrying about being looked down at:

\begin{quote}
``Because I know them so well I don't need to pretend like I am a upper level author that I know what I am doing all the time... I will just say to them like, 'hey guys I really can't solve this characterization right now.' It makes me feel more open to talk about it.'' (P9)
\end{quote}

Participants also explained that, oftentimes, they posted small pieces of their work in the hope of getting encouragement and affirmation. Their friends were likely to respond with praise and encouragement, which would be a great motivation booster to the writer. The social, friendly and playful atmosphere in group chats fed back to the effectiveness of the feedback exchange process. For example, P10 talked through an experience when they faced difficulties in writing and members in the group chat assisted in a relaxing way, helping them regain confidence in writing:

\begin{quote}
``They came in and kind of broke that tension in a really nice and organic way. And sometimes just, even though I'm very introverted, turning to to my trusted group of friends and having them help me troubleshoot is very, um, it turns a problem into something that's really fun and silly.'' (P10)
\end{quote}

\subsubsection{``Beta Reading''}
\label{findigns:private-feedback}

When writers finished a first draft or a large portion of the story, they commonly asked for feedback in the form of \textit{beta reading}, a widely used concept in the community that refers to feedback activities for a written but unpublished draft. Feedback providers, who were often called ``beta readers,'' would read through the entire draft, provide editing suggestions, and engage in discussion with the writer about the story content such as plot holes, characterization, and writing style:

\begin{quote}
``If I'm having problems after the writing process, like after I have the first draft and get something done, I sent it to what I call my `beta,' my editor, and I go: `read it over and make sure I haven't really screwed something somewhere. Make sure it looks okay and make sense.' I don't want myself to be like `oh wait somebody's arm was over here and now it's not.' '' (P3)
\end{quote}

%Beta reading feedback activities often happen nonpublicly between the writer and beta readers. Beta readers are often provided with unpublished drafts of the story, and have the permission to directly comment or even edit the drafts. Many of our participants also combine private communication channels. One common example is that the beta reader will read and comment on a shared document, while they chat with the writer in a separate synchronous chat tool, such as Tumblr or Discord direct message at the same time. Using this combination, the beta reader can leave major comments in the doc, while questions on small points, discussion and clarification can happen real time and back-and-forth in the chat. Similar to what we discovered in semi-public group chats, sometimes the conversation can divert from feedback exchange to general socialization and discussion on the context of the story. These type of socialization can strengthen the bond between the feedback provider and the writer, cultivating trust and consistent beta reading relationships.

Beta readers were provided with unpublished fanfiction drafts and given permission to directly comment or (in rare, high-trust cases) even edit the drafts. These interactions usually happened in private channels, such as privately shared text documents (e.g., Google Docs), one-to-one direct messaging chats, or emails. Coupled with the usage of private communication channels was commonly a close relationship between the writer and their beta readers. Participants described their beta readers as people whom they have an established trusting relationship. For instance, P3 reported that ``most of my beta readers are also my friends,'' and P1 summarized that their choice of beta readers were always ``someone whose opinions you trust, who you know.''

Close, one-to-one relationship was important to beta reading because it lowered social barriers associated with criticism. Participants indicated that established relationships made giving and receiving criticism easier, because it was considered more socially acceptable to ``be quite honest and say it's not working'' with a friend (P1). When the writer trusted the feedback provider, it was easier for the feedback provider to be completely honest and straightforward, because they would have had the expectation that the writer would receive their critique gracefully without worrying about harming the relationship: 

\begin{quote}
``If I have trust with someone already..., they can trust me not to take any feedback personally. I think a lot of people are reluctant to be overly critical of stories, [but] if you have that relationship, I know where they're coming from and they know where I'm coming from so they can say: `oh, I think you mean this or I think this might sound a little bit better.' '' (P10)
\end{quote} 

Established beta reading relationships also resulted in more substantial feedback exchange. Compared to in-progress feedback on snippets, where often writers are looking to share their excitement and get encouragement, writers more frequently desired serious, constructive feedback for their drafts before posting to public online spaces. For example, P12 mentioned that their beta reading was not ``just a simple exchange of a compliment and thank you.'' Instead, the feedback exchange involved ``more information shared'' and ``more communication going on.'' Writers trusted beta readers to provide informative and substantial feedback that would actually help them improve their work: 

\begin{quote}
``You also can trust that they're going to give you honest feedback. They're not just going to say, `oh yeah it's great,' and not actually point out all the gaping plot holes in it.'' (P26)
\end{quote}

%\begin{quote}
%It's building something with my friends. I know these people will do better than some of the people who leave a 'wow, I like that' sort of comment on a story, because there's just more information shared, there's more communication going on. It's not just a simple exchange of a compliment and thank you. It's not shallow. (p12)
%\end{quote}

Beyond being honest and pointing out areas of improvement, a beta reader as a friend would also be more thoughtful when providing feedback. Because the beta reader and the writer commonly socialized about life beyond fanfiction feedback, the beta reader had personal knowledge about the writer and the context of their writing. The beta reader would therefore tailor the delivery of the feedback to the writers' emotional status. For instance, P20 shared a story where they tried to help a friend's story with a critique, while and at the same time, communicated the critique in a considerate and encouraging manner given the writer's personal circumstance:

\begin{quote}
``I also know that for that particular person if I knew they were having trouble, they were going through something in their personal lives, I would kind of feel out if they weren't looking for that kind of feedback. I would make sure that whatever I said wouldn't add to their emotional distress.'' (P20)
\end{quote}

Relationships with beta readers also mean mutual knowledge about expertise and preferences. Asking for feedback from a stranger could be risky because, in P12's words, the writer might ``have no idea if this is something they actually want to read,'' and they ``don't want to throw it at someone if they're not going to like it in some way.'' (P12) On the contrary, when working with beta readers that they were familiar with, the writer could selectively ask for feedback that aligned with their strengths and interests:

%\begin{quote}
%"My beta varies in different fandoms depending on which one I was writing in... I tend to reach out to them based on who betas what type of fic for me. When I write for \textit{Yuri! On Ice} is that I have a certain person that I go to and she reads all of those fics. When I work in the \textit{Hockey} fandom, I have a friend who reads all of that stuff I've got to try to work at. I got a friend, when I wrote my K-pop stuff, she's the only person who touches all of that... [It depends on] how much can they contribute, how much they already institutionally know about this thing." (P3)
%\end{quote}

\begin{quote}
``If it's a certain type of story, I'll contact someone who I think might be more interested, you know, certain fandoms, obviously, certain ships, I'll ask different people to take a look at.'' (P10)
\end{quote}

On the other hand, feedback providers who knew the writer well often gave feedback that was tailored to the writer's strengths and weakness, writing style, and usual feedback needs. In many cases, they would have a shared context about the story, which made the feedback exchange process more efficient. In the words of the participants, a beta reader is someone who ``understands what I'm going for usually,'' (P10) who they ``have been writing for a long time with'' so they ``trust their judgement'' (P29), and who ``will be looking for sorts of things that I'm looking for them to look for.'' (P26)

Lastly, a close relationships with a beta reader lowered the barrier to asking for feedback that required extensive effort. Beta reading usually takes a lot of time and effort, especially for long and multi-chapter fics. Writers preferred a beta reader that stuck through the whole story in comparison with someone who jumped in once in a while, because of the substantial investment needed to get familiar with the deep content: 

\begin{quote}
``It's easier because if they've already read the first five chapters they can just read six, seven, and eight. Whereas for a new beta reader, all of a sudden they have to read not just the three new ones but also all the past ones. It's easier to say `hey I'm going to write this 45 chapter fanfiction,' than `hey can you read this 45 chapter fanfiction...' '' (p24)
\end{quote}

With an established relationship, it was also more acceptable to follow up with a question and further discussion on the feedback: 

\begin{quote}
``[Because of] the fact that we're good friends and we have that relationship, I felt okay going back to her and asking for clarification about that scene.'' (P10)
\end{quote}

\subsubsection{``All I Want for Christmas are Reviews''}
%%Public (FFN, AO3, social media)
%%%Shallow encouragement
%%%Specific constructive (rare)
%%%How: Culture / norms of commenting

After publicly posting their stories online, writers got feedback in the form of public reviews, a common affordance of fanfiction websites and social media. Although this type of feedback has been studied the most in research literature, it is in fact the last stage of feedback exchange in the creative process. In contrast to the earlier stages, the most common type of public review, as observed by our participants, was simple and surface level words of encouragement from online strangers. Although not directly helpful to their story and writing skills, encouraging words and positive reactions provided writers with indicators that their genre and plot worked for their audiences, which boosted their confidence and incentivized them to write more in the future.

In order to get more reviews, writers strived to ensure that their fics would be broadly visible to an interested audience. Participants reported a strategy of ``cross-posting,'' meaning they would post their fanfiction on multiple platforms, ranging from sites specialized for fanfiction archival such as FFN (Fanfiction.net) and AO3 (Archive of Our Own) to general social media platforms such as Twitter and Tumblr. As P10 pointed out, ``If I'm posting something on Archive of Our Own, there's always a post on Tumblr about it.'' Cross-posting helped ensure the fanfiction reached a wider audience than posting on a single platform.

Participants shared that they would tailor their posting style to the different user preferences and behaviors on different platforms. On the one hand, fanfiction sites had an audience that was genuinely enthusiastic about the story itself. For this reason, readers on these platforms were considered to be more ``serious''\textemdash as P13 described using imagery, readers on fanfiction sites would read fanfiction like they were ``reading a book by fire in the evening.'' As a result, writers would post well-polished, sometimes long and multi-chapter fictions on those platforms. On the other hand, posting on social media would ensure more serendipitous exposure of the fanfiction, reaching a larger body of readers who were not yet aware of the particular story. Therefore, writers posted their fictions on social media as an advertisement. Sometimes writers would ask their followers to share their social media previews, broadcasting the fiction to an audience outside of their followers. Besides increasing general visibility, writers also considered it important to make their stories seen by readers who would truly appreciate their stories. 

Posting on a platform which afforded tagging and categorizing stories was considered an important factor for this purpose. On both social media and fanfiction archival sites, writers would spend a lot of time and effort on tagging, with the hope that the fiction would be discovered easily by the target audience. They would also leverage the tagging system to mark out potential triggers, content that might disturb some readers or reraise trauma. Trigger warnings reduced the risk of receiving comments attacking the story and writer from someone upset by parts of the content. 

Another practice for soliciting public feedback was to communicate appreciation for reviews with readers. Participants reported that in ``author's notes''\textemdash a common fanfiction practice where writers attach a short block of text to the beginning or the end of their fiction\textemdash they would explicitly ask for reviews, for example: ``All I want for Christmas are reviews'' (P2), ``positive reviews are welcomed'' (P4).

Although participants generally appreciated any encouraging reviews, they preferred reviews that were both positive and constructive, providing advice for improvement and ideas for new stories. In order to get more positive and meaningfully constructive reviews, writers made sure that their fanfiction reached the right audience who would appreciate the story. Writers chose platforms based on where they believed their audience would prefer to see the content. For example, a story about a certain pairing had more supporters on FFN compared to AO3, so writers would post the story on FFN in order to get more praise and avoid conflict. Another participant described AO3 as a site with a more inclusive culture, so they would be more likely to post queer-related stories on AO3 than on FFN. Writers learned the culture of different platforms by observing the amount of fanfiction about a specific topic/pairing and the amount of positive reviews.

\subsection{How do fanfiction writers identify affinity and build online connections for feedback?}
\label{finding:relationship}
For writers who do not already have established feedback connections, it is important to build new connections, especially with those in the fandom who have the interest and expertise to be feedback providers. However, finding new connections in the community is never easy or guaranteed. In this section, we report our findings on five practices that fanfiction writers used across public and private channels to find and build feedback relationships: connecting through public comments, participating in social events, ``just-reaching-out,'' engaging in small private communities, and disclosing identities.
%Finding/initiating a connection (Just reach out)
    %[x] Replying to all Public reviews 
    %[ ] Community activities
        % events
        % collaborative writing/role playing/chat ficing
    %[x] just reach out

%Deepening relationships
    %[x] Identity affinity & self-disclosure
        %IRL Identity disclosure
    %[ ] Socialization beyond feedback
        % chatting and giving support for real life stuff
    %[ ] small close community

\subsubsection{Connecting Through Public Comments}

%Getting exposure is crucial to getting feedback. Fanfiction writers, after finishing and posting their works on Archive of Our Own, would use a number of strategies to try and get more exposure on their works, in an effort to attract an audience who would connect with their work. Crossposting, tagging, and participating in social media by following, reblogging and linking were all strategies to increase exposure: 
Connections with feedback providers often started in public spaces. The public comment sections under published stories in fanfiction archival websites such as FFN and AO3 were a salient place for writers to identify feedback providers. One important strategy was connecting through public comments. Although comments were often a valuable source of feedback in-and-of themselves, each new comment could also function as a digital introduction to someone in the community. Several participants talked about relationships that initiated through comments:

\begin{quote}
``I've made friends with a lot of people who started out just commenting on my fics a lot. You end up commenting back, and start talking... It's like hand picking your friends, this person already writes all these things.'' (P13)
\end{quote}

Writers noticed people who were repeatedly commented on their work. ``You start recognizing the tag sign and the pictures'' (P28). That name recognition from repeat commenting was sometimes a starting point for building relationships across platforms:

\begin{quote}
``I might go and read their work and leave a comment, or they might end up being part of the same discord, or if I do come across their Tumblr I might look at it. I usually respond to the comment and say `thank you,' and if they left analysis, I (will) talk back and forth.'' (P29)
\end{quote}

One strategy writers used to make new connections was to comment on the fics they were reading, making sure to comment after each new chapter was posted. By repeatedly commenting on the fics they were reading, writers started building name recognition with others in the community: %This was also part of a reciprocal norm of commenting.

\begin{quote}
``I try to treat people how I would want. If I read a fic that is good, I leave a message, saying hey, here's what I liked, keep up the good work. If someone takes the time to actually communicate, and I'll leave one on every chapter. I've actually as an author, always reply to comments. '' (P7)
\end{quote}

Additionally, writers would reply to every single comment they received, which was often considered as a reciprocal norm in the community:

\begin{quote}
``You have regular commenters who read all your stuff or subscribe so they get alerts when you post something new. I try to reply to every comment that I get and I think it encourages people to respond more... I think I get more comments if I comment back.'' (P13)
\end{quote}

As P13 stated, replying to comments was considered to be a useful strategy to solicit future public feedback. This widespread norm helped build connections and establish a sense of community. However, P13 also pointed out that in the case of larger fandoms, they would get ``a ton of comments'' and they would ``not really expected to reply since it's such a huge volume.'' Therefore, it was relatively more difficult to make one's comment stand out or spot salient commenters in large fandoms. In other words, initiating feedback connection through public comments was considered more challenging in this situation.

\subsubsection{Participating in Social Events}
Participating in online fandom events was another way of initiating and building connections for feedback. These events were organized by fans in a variety of forms and scales, combining fanfiction with fanart, role play, discussion forums, video games, etc. Events encouraged writers to write and share stories based on certain topics, creating spaces where they found new connections with people who had similar interests in the topic, story, or even specific relationships between characters. During the events, writers often had opportunities to interact with each other and engage in each others' stories. In many cases, the connections they made in these spaces turned into online friendships and feedback providers. For example, P29 shared the story of meeting both of her closest beta readers in fandom events:

\begin{quote}
``[In the case of her first beta reader], we joined the same forum for girls into star wars. We shared stories with each other and helped each other on stories. [That was] November 2011 [when] we were doing Nanowrimo (National Novel Writing Month). And the other friend, I met her in the youth forums in the roleplay section. There was this roleplay, everyone made their own character, we got to know each other through writing together on that roleplay.'' (P29)
\end{quote}

%For instance, one participant told us about a long-term relationship that began within a fandom subreddit. Another participant talked about how friends they made in world of warcraft became feedback providers. And another participant talked about meeting new audiences in while posting fan art to tumblr. Although these spaces are not fanfiction specific, writers are able to find people with a shared affinity for fandom, and leverage this affinity to make connections.

Such community writing events were usually organized with the goal of facilitating opportunities for people to connect. P2 talked about how they organized a Secret Santa, a holiday exchange where members of the fandom submitted prompts for a story they would desire to read. Event attendees swapped prompts and wrote each other fanfiction. This provided attendees with an opportunity to connect with someone who shared interests. In addition, the organizers of these events often required attendees to have their work go through beta reading, and volunteers would be there to beta read the works. Our participants shared their appreciation of these events in that they offered chances to connect with and test out a new beta reader, getting feedback and making a new connection simultaneously. For instance, P20 shared that normally, they often had difficulty finding beta readers online because they were unsure about whether strangers would be interested in and committed to their stories:

\begin{quote}
``Finding betas for my fics is always super hard,... I know for longer fics or works in progress especially, it's hard to find someone who's committed to, not only taking on a longer project, but also just kind of committing to whatever content I was writing about...''
\end{quote}

\noindent For P20, participating in fandom events was an effective way to not only get exposure to a group of writers in the community who were interested in beta reading for others, but also to learn about their strengths and interests so that they could identify suitable feedback providers for their work: 

\begin{quote}
``The times I did find beta readers and they were committed to what I was doing would be like for a challenge, like a `big bang,'\footnote{A kind of fan community event where writers write and share stories with assigned artists in the fandom, then the artists create fan art work based on the stories. The ``reverse bang'' mentioned next is a similar community event, where writers write fanfictions based on fan art works. More information about these fandom events can be found here: \url{https://fanlore.org/wiki/Big_Bang}} or a `reverse bang,' where it was required that every fic was beta read by someone. So if I was writing for a `big bang,' either the moderators would assign a beta reader to me, or they would have a list of beta readers who were interested in beta reading, and I would literally just go down the list and find someone. From my experience, the way those lists were formed [is that] the beta reader would list their strengths.'' (P20)
\end{quote}

%They would also list their weakness and all that stuff... I would kind of cross out the ones that couldn't't beta my content because they either, their strengths were for something I was not looking for, or they listed a trigger that was inside my content and that just cut down my list significantly. When I went through that list, and I had a few people in mind to contact, I would actually try to find either like their writing profile to see if the'd written things before, or if they had a blog, literally I would just read their about and see if they used correct grammar and English. So that would help out a lot. p20

As we mentioned in Section \ref{findigns:private-feedback}, writers found it challenging socially to reach out to an online stranger, even in cases where they have identified the stranger as a potential good feedback provider. Community events offered both the communication channel and ``excuses'' for writers who would like to reach out. For instance, P9 shared their experience serving as a beta reader in an event:  

\begin{quote}
``I have two Discord servers and one of them was just for this one ship in [my fandom]. And its just for writers, fan artists, and `scholars' as the title of the role, which are the beta readers... You can pin the scholars on Discord or add all the scholars. So they will be like `hey, I need a beta reader, can someone look at this?' Then I will get the notification so I can look at it if they want me to.'' (P9)
\end{quote}

In the community event group chat, because of the same interest and the sense of community created by events, writers often feel the easiness to externalize their their need for feedback and reach out: 

\begin{quote}
``We did a `big bang' in the fandom and in the side bar within a channel, there was a person saying `who can help me decide a major plot point?' And I was there, I said that I can help you. Then we started to chat that. '' (P11)
\end{quote}

\subsubsection{``Just-Reaching-Out''}
When we asked fanfiction writers for their advice to newer writers in seeking feedback, the almost unanimous recommendation was to ``just reach out.'' This advice reflects the most common barrier fanfiction writers face in making connections: a feeling of social anxiety. Writers described the fear that the person they were reaching out to could be too busy, or unfriendly to cold reach-outs. But participants who provided this advice said they had experienced relief to find that the other writers they reached out to were responsive and positive.

Writers recommended asking for feedback in cold reach-outs via social media posts, comments, group chat messages and private direct messages. In a social media post, a writer might describe their work and state what they're looking for in a feedback provider. Then they would reach out to people who comment on the post volunteering to give feedback (P26). Another participant shared an experience of asking for help by commenting on a favorite author's post:

\begin{quote}
``That's how I got started. I made a comment on a story on Fanfiction.net that day: `I've got this story I want to write, what should I do?' And they PMed (private messaged) me and told me exactly what to do. And they encouraged me to write it, they encouraged me to put it up and I did. And there was just no stopping me after that.'' (P6).
\end{quote}

In addition to public outreach, writers recommended reaching out to people in direct messages. One strategy was to find a group of people and reach out in a group chat:

\begin{quote}
``I would say the most reliable in terms of the people who keep coming back over and over, even as I switched fandoms, it's been finding friendship group first. Find a group of people who tend to agree on your weird fandoms and your divisive opinions and create that community, and then show them your precious little child and say, what do you think?'' (P23)
\end{quote}

An out-of-the-blue direct message was a tried-and-true strategy among several participants. Several writers expressed the sentiment that social anxiety is a widespread problem holding them back, but after having reached out, they've been relieved to find that the community is overwhelmingly kind and helpful:

\begin{quote}
``I know 90 percent of the reason most of us are on this site is we're introverts who have anxiety. We're really scared of putting ourselves out there, but if you just ask someone, if you just talk to someone, you know, make a post and share it around, I told someone the other day that most writers would rather die than discourage another person writing. So even if they don't have time or they don't think they're the right person to read this thing, so few people will be outright rude or cruel, and so everything that all of us are afraid of is kind of silly [laughs]. I've just found every single time I've reached out to someone in fandom they've reached right back in their own way. So I'd definitely say, that is the advice is just go for it.'' (P15)
\end{quote}

%\begin{quote}
%“Fandom is very welcoming and it's very like ‘take care of our own.' We can sort of come in, sit down and have a drink, we'll take care of you and we'll help you however we can. And I've found that to be true across multiple fandoms and over the years.” (P14).
%\end{quote}

Whatever the channel was, fanfiction writers found a lot of success in making feedback connections by just reaching out to others in the community.

\subsubsection{Engaging in Small Private Communities}
%you're talking to the person and get a feeling of that person. I think it took me a good six months to disclose I had children. I think that was my main concern. Because obviously you went to protect your children's identity. But I think after a year we both kind of knew everything inside out about each other because [inaucible]. And the fact is I had known the people had kind of vouched for me at that time. So she was more confident in knowing, who I was (p28)

%Actually one of my friends who's also one of my beta readers, we met through Tumblr and we learned sort of like we can miserate together over our grad school experience. That's how we built this connection. (p3)

Beyond close one-to-one interpersonal connections, fanfiction writers found connections with small, close communities, friend groups where they felt comfortable getting ideation and in-progress feedback. A small, close community could take the form of a Discord server, a chat group on Facebook or Skype, a board on a less-traveled forum, or the right intersection of tags on AO3. These channels were characterized by being highly niche, comfortable, and small enough in scale that everyone participating could get to know each other. Earlier, we discussed the role of these highly trusted groups in feedback, but these small, close communities also provided supportive connections writers needed to thrive. For participants, this meant connecting with relatable people who made them feel comfortable, encouraged them, and gave them feedback on ideas and writing. For example, one author spoke with us about a set of Discord servers that brought together queer women who shared interest in a certain pair of characters from a certain animated series:

\begin{quote}
``I particularly like that there's sort of these little communities of queer women or mostly queer women or queer aligned groups… It's just nice to talk to people who get it, who get why you're so excited.'' (P4)
\end{quote}

\noindent The people in this group had a common ground because they shared an underrepresented identity and they were into the same fandom. We define the term \textit{affinity intersections} to describe the small, close communities occurring at niche intersections of interests and identities. In P4's experience, affinity intersections created a safe environment to talk about writing queer sexuality into fanfiction:

\begin{quote}
``I've seen how friendly and nonjudgmental everyone is in responding [to others]. That makes me feel quite safe to go and ask them, `how do I write this thing?' And it's something that's quite sort of deeply personal and intimate.'' (P4)
\end{quote}

A shared Discord server provided writers with a safe place where they could connect with each other in a carefully moderated and curated group. Chat groups also became spaces to commiserate, give each other encouragement, and hold each other accountable for writing. This did a great deal to help writers break through and make progress when they were feeling frustrated or stuck. Writers organized little `sprint' events, where they each agreed to write as much as they could for a short period of time:

\begin{quote}
``We will set time and be like `in the next 30 minutes, we are going to write as much as you can and when we come back, share the sentences...' Some people come back be like `I wrote a thousand words' and I will be like `I got 10'. I will be like I didn't come up with anything but they will be like `well those 10 words you didn't have them before.' So overall it's a positive thing.'' (P3)
\end{quote}

Connecting with a small group at an affinity intersection was also a great way to meet feedback providers. Since these small communities were places where people shared the same niche interests, writers felt that there was a high likelihood that others respond to requests for feedback. Having an ongoing relationship with feedback providers helped writers get deeper, more thoughtful feedback. They felt understood by their in-group because of their shared context.

\begin{quote}
``They've all read my fic pretty in depth. So I can be like remember when this happened, or where should I go for this part of, you know, my next venture into this universe or whatever. They know what's up there so I don't have to re-explain everything or force them to watch the show or something, so they can understand what I'm thinking all the time.'' (P17)
\end{quote}

The benefit of affinity intersections boils down to being understood by others. Writers in these tight-knit communities mutually understand each others' interests, their writing contexts, and the experience of writing fanfiction. As a result, writers in the same close community would offer encouragement and comments to each other in public spaces like Tumblr and AO3.

\begin{quote}
``When I get the same people commenting on things that I've written, that makes me feel like I'm part of a little group... I'm part of the gang that does this. And privately talking to people who's stuff I read who are other fans, it's a quite nice feeling of belonging... there's a sort of comment exchanging between writers in fanfic, you know, I'll comment on yours and you'll comment on mine, cause we all know how much we love it.'' (P4)
\end{quote}

To summarize, small, close communities formed around shared interests, creating spaces where people felt comfortable, found connections and received support. People maintained the small community connection in public internet spaces, helping promote a sense of support even while participating in larger fandom contexts. This was particularly important for fanfiction writers that tackled topics that were underrepresented, taboo, or centered on marginalized identities.

\subsubsection{Disclosing Identities}
%%Low trust on online stranger
Finally, disclosure of IRL identity was a step that some fanfiction writers took to different degrees in their online relationships. With only one exception, all participants posted their fictions pseudonymously, and therefore, fanfiction community members knew each other primarily by their internet handles. IRL identity disclosure could be an incidental or intentional step as writers exchanged feedback and built relationships with each other. Some participants discussed occasions where they disclosed their IRL identity to people whom they had initially met through fandom: 

\begin{quote}
``Once you get past that little bit of a hump of being almost like unsure if you should identify yourself or not, you can make some really good friends. And it's been, it's an honor when somebody says: `oh, by the way, my name is.' And then of course you say: `well, that's great. My name is,' and hopefully you can build friendship.” (P6)
\end{quote}

Several writers connected IRL identity disclosure with building friendship and trust. A norm in place was that writers did not ask people to disclose their identities, so much as take the step themselves. If the other person reciprocated, this would change the dynamic of the relationship to be closer:

\begin{quote}
``I'm not calling them by their username in my head anymore. And a person we've been messaging for 5 years, we learned each other's name 3 weeks ago. It's a warm fuzzy feeling, and we message now more than we used to. I think it definitely is part of [becoming closer].'' (P15)
\end{quote}

However, writers noted a barrier to IRL identity disclosure: a culturally widespread fear that internet strangers are dangerous. When talking about disclosure, writers named themselves, their siblings, children, and close family members as people they wanted to protect by remaining pseudonymous:

\begin{quote}
``I think like a lot of people in who are now in their twenties, I grew up on, you don't give out your real name, you don't say what city you're in. There are mad axe murderers on the internet and they will track you down.'' (P23)
\end{quote}

In their experiences with actually disclosing their identities to other individuals, our participants dispelled this belief, noting that most fandom people were actually nice. This fear of real-world violence created a barrier to connection with others online, and in overcoming this, they found rewarding relationships. However, participants shared that basic precautions and common sense were still important for deciding when and to whom someone should disclose:

\begin{quote}
``There's not any set guidelines. I think it really depends on who you talk to... how long have you been with the person? What type of things do you talk about? Do you feel like it's safe to give that information? ...You kind of have to sometimes make a snap judgment and ultimately it worked out fine in this one case... it really does have to come down to instinct, gut, sometimes, there's no kind of set formula to be sure.'' (P21)
\end{quote}

In the cases where a trusting relationship fostered by identity disclosure was established, such a relationship is in fact intertwined with feedback relationships. Participants shared that close relationships with beta readers were at times the same relationship with IRL identity disclosure. For example, P11 shared that: 

\begin{quote}
``And also since me and my betas work on [Google] drive, and the drive documents are linked to my personal email, they can see my name, but it's a rare occasion. Few people know my name and surname that in the fandom community.'' (P11)
\end{quote}

Although real life identity disclosure was not strictly necessary for critique, the type of trust needed for identity disclosure was transferred to trust in the sense of believing someone will give well-intentioned, constructive, and accurate feedback.

%This raises a question: is there some transference, whereby trusting that someone will not misuse your personal information also builds trust that someone will provide the right kind of feedback?
%%Newcomers 
%%Awkward to join already connected community 
%%Expectation on effort 
%%Maintaining feedback and beyond relationship
%%Technological adoption
%%Social anxiety for reaching out
%%Don't know where to find expertise -> lack of affordances for discovering the right people
%%(Getting Exposure) Not receiving meaningful public reviews or attention 

\section{Discussion}
%Different stages and types of feedback (v.s. one-size-for-all existing feedback systems)

%Importance of connection and affinity in feedback (v.s. low emphasis on connection in current feedback systems)
%%How self-disclosure ->mutual trust -> feedback
%% trust makes feedback relationship different

%How does this deepen our understanding of online affinity? 
%Intersection of identity affinity and interests
%%Navigating hierarchical affinity spaces

It is no secret to the CSCW community that creative work benefits from collaborative effort. Taking a different approach from the rich literature on creativity support via crowdsourcing, our study further unpacks how creativity and feedback exchange can thrive in a socially situated and personally interconnected manner. In this paper, we presented findings from our interview study with 29 online fanfiction writers, surfacing their feedback practices along with how they identified and built feedback connections in online affinity networks. Next, using the example of fanfiction writers, we synthesize our findings into theories of creative feedback exchange: how creators' social needs in feedback vary feedback during different stages of the creative processes (§\ref{discussion:social-needs}), how feedback and personal relationship development are intertwined, (§\ref{discussion:feedback-affinity}), and how feedback exchange crosses an ecology of online platforms (§\ref{discussion:ecology}). Finally, we discuss the implications to the design of future online feedback systems (§\ref{discussion:implications}).

\subsection{Social needs in feedback exchange throughout the creative process}
\label{discussion:social-needs}

\begin{figure*}[h]
  \centering
  \includegraphics[width=0.7\linewidth]{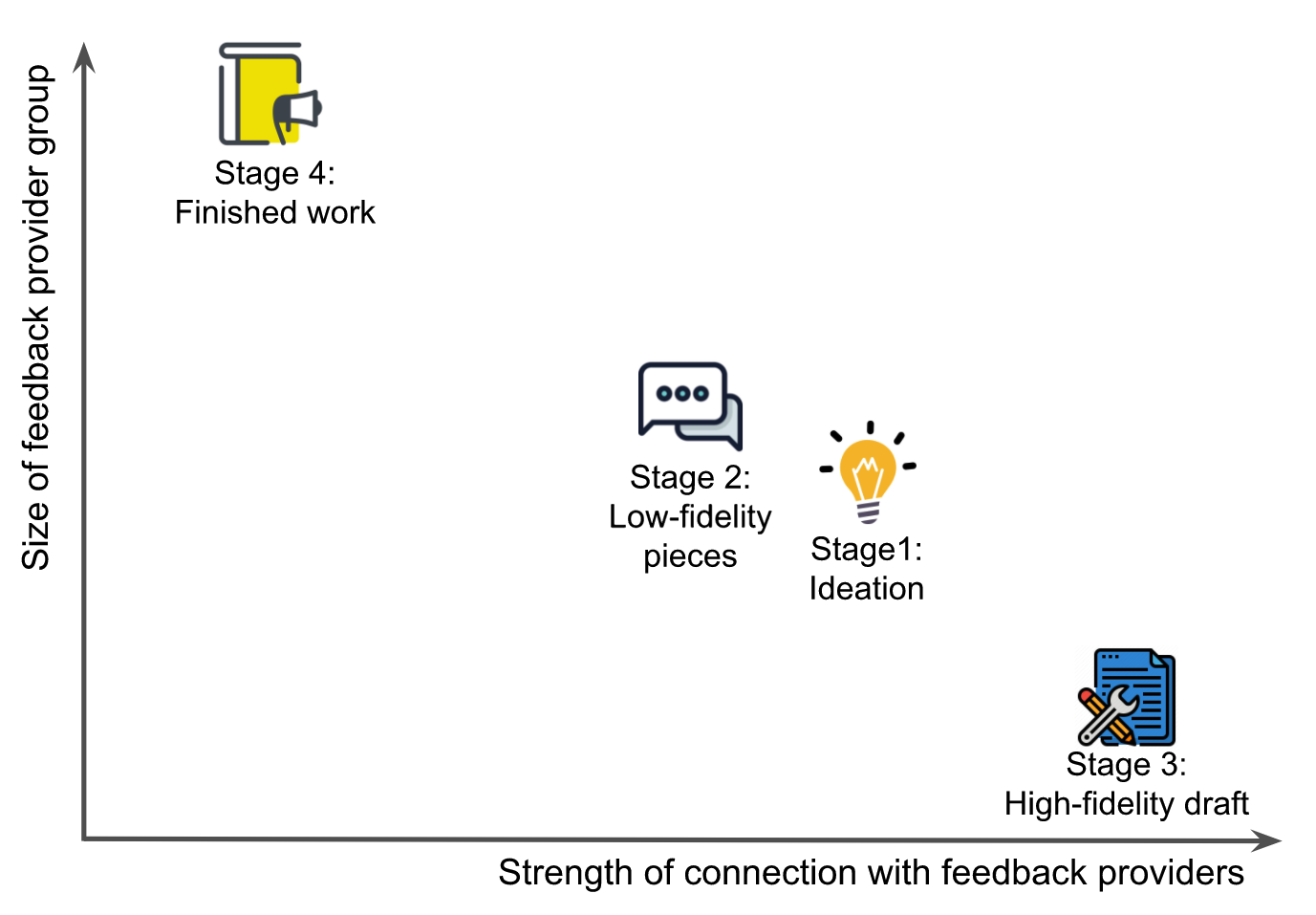}
  \caption{Needs in the size of feedback provider group and the strength of connection with feedback providers in different stages of creative process. }
  \Description[Feedback Provider Group Size and Connection Strength across Creative Stages]{Icons represent four creative stages on an X-axis, titled “Strength of connection with feedback providers,” and Y-axis titled “Size of feedback provider group.” Stage 1: Ideation is placed with high strength of connection and low size of group. Stage 2: Low-fidelity pieces is placed at a slightly lower strength of connection and slightly higher size of feedback group. Stage 3: High-fidelity draft is placed at the highest strength of connection and lowest size of feedback group. Stage 4: Finished work is placed at the lowest strength of connection and highest size of feedback group.
}
  \label{fig:social_needs}
\end{figure*}

%From R1: this is a glimpse into private online spaces and that this sort of exchange may happen between friends elsewhere. As such, these interviews offer an interesting glimpse into how close-knit communities give and solicit feedback in small, private spaces.

Creative work involves multiple stages, such as brainstorming, implementation, iteration, and presentation, while the need for feedback is presented at all stages. Furthermore, as \citet{foong2017online} pointed out, online feedback exchange is a complex multi-step process. Deriving from our specific example of fanfiction writing, our results add a new dimension to \citet{foong2017online}'s framework of online feedback exchange\textemdash the different \textit{types} of feedback sought out and exchanged at corresponding stages of the creative process.

In §\ref{finding:feedback-practice}, we described four distinct types of feedback practices in the words of our participants: \textit{``throw ideas at each other''}, \textit{``give my friends a snippet''}, \textit{``beta reading''}, and \textit{``All I want for Christmas are reviews''}. Adapting \citet{amabile1988model}'s model on the creative process, we map each feedback practice to a stage in the creative process. In the \textit{ideation} stage, creators brainstorm together and exchange inputs on each others' ideas; In the next stage, creators have done some initial generation and collected some \textit{low-fidelity pieces}. In our example of fanfiction writers, writers show each other snippets of work and hope for reactions. After having a \textit{high-fidelity draft}, creators look for feedback for the entire draft and details. In the case of fanfiction writers, that means intense, top-to-bottom ``beta reading.'' In the final stage, when the creators have finished their work, they present it to the public for feedback. 

Our study uncovers distinct social needs in feedback exchange for these four stages. While studies of online feedback exchange laud the quantity of feedback and providers, resulting in systems that aim for collecting feedback at scale (e.g., \cite{xu2014voyant, luther2015structuring}), we discovered a much more intricate story from fanfiction writers\textemdash a multi-layer social network that includes many shallow, infrequent reviewers and a few close, substantive feedback providers \cite{davis2021mentorship}. In Figure \ref{fig:social_needs}, we map out the social needs in terms of the need of the strength of social connection with feedback providers and the need of the size of the feedback provider group across different spaces. We explain the needs as follows using the example of fanfiction writers. 

In Stage 1, creators often need feedback providers to spend time with them to ``throw ideas at each other.'' While most research focuses on public online spaces, where feedback often happens unidirectionally and asynchronously between online strangers \cite{kou2017supporting, evans2017more}, we surfaced feedback practices as early as when creators are still forming their ideas. In the case of fanfiction writers, this type of feedback exchange is interactive, playful, and socially embedded. Therefore, instead of posting their ideas in public, creators engage in this kind of feedback interaction with a small group of people who are already friends with them and in relatively private spaces, where they can chat in real time and have a dynamic conversation. 

Similarly, in Stage 2, writers write small, experimental snippets of their story and gradually build up a bigger plot. Echoing prior literature \cite{crain2017share, kotturi2019designers}, our findings show that creators in general consider such low-fidelity works-in-progress as informal and private. They refrain from presenting low-fidelity work for feedback in completely public spaces. However, contrary to assumptions in previous literature that creators generally do not spontaneously seek feedback online for early in-progress work \cite{kim2017mosaic, kotturi2019designers}, we found that they indeed do, but instead in relatively private channels among people with whom they have an ongoing connection. Despite the needs of privacy and close connection with providers, feedback from multiple perspectives is welcome, as creators would like to test out pieces of work, collect reactions, and avoid any potential problems at an early stage. In our case study of fanfiction writers, they also encountered creative blocks in the process of writing the story and sought encouragement from friends. 

In Stage 3, where creators have put together a high-fidelity draft of their work, such as a completed first draft of a fiction, creators need one or just a few consistent feedback providers with whom they have a close relationship. Previous literature has shown that after receiving feedback, creators need to sort out useful pieces of information from feedback that involves multiple or even contradictory suggestions \cite{dow2013pilot}. This can be challenging when they are unfamiliar with the feedback providers' expertise and perspectives \cite{dow2013pilot, connor2015discussing}. Fanfiction writers avoid this issue by seeking feedback from a single provider who has worked with them previously, so that they are familiar with their style, strengths, and needs. A close relationship with their feedback providers also means that providers are more likely to invest the effort needed to generate thoughtful and constructive feedback. 

Finally, in Stage 4, creators post finished work on social media and online archives. At this stage, creators are showcasing their work, and they appreciate or even adore comments coming in from a large number of online acquaintances and strangers. Prior studies of online mentoring have identified types of feedback in public online spaces, including encouragement, constructive critique, and discussion-style sense-making interactions \cite{kou2017supporting, evans2017more}. Our findings add that, at this final stage of creation, creators want to seek affirmation for their work and develop the confidence to continue future creation. Criticisms, although in some cases may be constructive, are frequently deemed as inappropriate and in fact relatively rare \cite{evans2017more}. We also uncovered that creators use public feedback as a channel to identify potential new feedback providers who hint their expertise in public comments by complimenting specific aspects and excerpts in the writing. 

%including problem or task identification, preparation (gathering and reactivating relevant information and resources), response generation (seeking and producing potential responses), and response validation and communication \cite{amabile1988model}, where feedback can come in at each of the stage. 

%This case study of fanfiction contributes a greater understanding of distinct feedback needs at different stages in creative work and affordances needed to the broader CSCW literature on online feedback.

\subsection{Intertwined development of feedback and personal relationships: from public spaces to private spaces}
\label{discussion:feedback-affinity}
%The role of affinity is to motivate social connection and learning, and to situate knowledge as social capital within the community. Similarly to how Lave \& Wenger theorized mentoring and learning in their foundational work on \textit{Communities of Practice} \cite{lave1991situated}, learning in online affinity networks is theorized by Ito as the adoption of community practices, rooted in a sociocultural learning perspective. In other words, the skills learned are not easily separable from the social context in which they are learned. However, unlike communities of practice, online affinity networks do not have strongly formalized roles, relationships and institutions. Instead, roles in affinity spaces are fluid and dependent on expertise within domains related to the affinity, so that someone could be an expert in one domain and novice in another \cite{ito2018affinity}. This also creates a challenge for learners: rather than relying on a formal institution to ascribe members as mentors and experts, learners must identify and develop relationships with people of desirable expertise. An examination of feedback-seeking within the model of online affinity networks can drive insights around how individuals build and leverage their personal affinity network across affinity spaces for the purpose of getting creative feedback.

\begin{figure*}[h]
  \centering
  \includegraphics[width=0.8\linewidth]{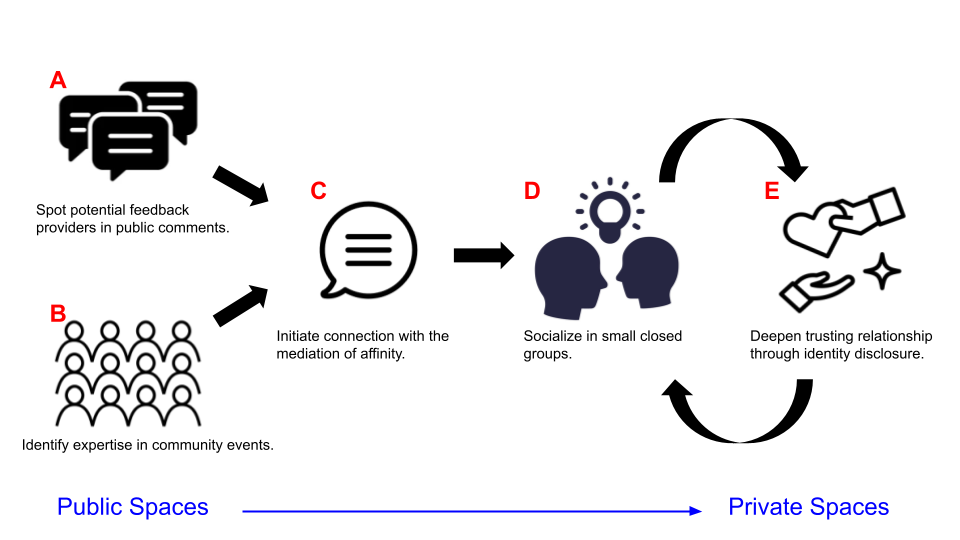}
  \caption{A hypothetical user journey of finding and forming relationships with feedback providers in online affinity networks.}
  \Description[Hypothetical Journey of Developing Feedback Relationships]{Five stages of relationship formation are placed on an X-axis starting with public spaces and moving towards private spaces. The first two stages, “spot potential feedback providers in public comments” and “identify expertise in community events,” are displayed in parallel. Arrows from these first two point to the third stage, “Initiate connection with the mediation of affinity. Another arrow from there points to “Socialize in small closed groups.” Lastly, there is a bidirectional connection from this fourth stage to the fifth, “Deepen trusting relationship through identity disclosure.”}
  \label{fig:feedback-relationship-journey}
\end{figure*}

Our study points to the conclusion that personal, authentic connection with feedback providers is crucial to effective feedback exchange. Moreover, prior research observes that it is challenging for creators to find high quality, stable sources of feedback online \cite{xu2012what,hui2019distributed}, especially for newcomers \cite{marlow2014rookie}. Our case study illustrates that the feedback exchange in public fanfiction archives is not just ``feedback exchange'' per se, but also an integral part of finding new connections with people who share affinities. In §\ref{finding:relationship}, we report our findings about how fanfiction writers identify and build relationships with feedback providers. We summarize this process into a hypothetical user journey presented in Figure \ref{fig:feedback-relationship-journey}.

We explain the development of feedback relationships using the example of fanfiction writers. For a new writer who has just started sharing their stories online, identifying potential feedback providers in public feedback exchange spaces (i.e., comment sections on fanfiction websites) is a good first step towards building relationships for feedback, as indicated as step A in Figure \ref{fig:feedback-relationship-journey}. Strategies include commenting on every fic they read, as well as replying to every comment they received. Writers who comment on everything they read do so in order to respect for the culture of the space by reciprocating others' efforts. This finding complements a prior study of a photography critique community, where reciprocal feedback exchange was desired but not often observed \cite{xu2012what}. We add that exchanging public comments in a reciprocal manner also means exchanging information about interest, strength, and expertise, offering a pathway for creators to forge new connections with their audiences.

An alternative approach to initiating a relationship for feedback is through participating in community social events, as shown as step B in Figure \ref{fig:feedback-relationship-journey}. In contrast to peer- or crowd-based feedback systems where the exchange is completely separated from socialization, our finding suggests that feedback exchange can be enhanced by the opportunity to socialize with feedback providers. As described by our participants, certain features of fandom events (e.g., a list of beta readers) helped them identify who would have the interest and expertise needed to meet their feedback needs. Echoing prior literature on how online participatory culture fuel learning \cite{jenkins2006convergence, black2009online}, our finding indicates that a social atmosphere created by shared interests and passion can support effective feedback exchange.

%Although our study examined fanfiction and theirs photography, both sets of interviews point to the design implication of instilling reciprocity into the culture of the community. 

%As we saw, many of the interviewees' most important feedback relationships started with the comment exchange. Thus, although reciprocal public feedback exchange is important for affirmative and constructive feedback, its utility for making new connections should not be overlooked. Public feedback exchange should be considered as part of a holistic process that includes both feedback exchange and connection building.

%One theme that continued throughout this study was the interrelatedness of feedback seeking and connection building. Throughout the process of writing fanfiction, from ideation to editing to posting, feedback seeking and relationship building co-occurred, with each influencing the other in turn. 
%Likewise, fanfiction writers initiated and developed connections across public, semipublic and private spaces. 
Practices of reciprocal public feedback exchange and participation in social events helped lower the social barrier to initiating feedback request (step C in Figure \ref{fig:feedback-relationship-journey}), as it is mediated by the affinity and rapport built in the previous steps. Our participants advised newer writers to ``just-reach-out'' to established members in fandom for feedback, which seems contradictory to previous research that indicates novices tend to reserve to themselves in terms of feedback because they do not want to expose their vulnerable status \cite{marlow2014rookie, cheng_critique_2020}. Our explanation is that shared interests and identities in online affinity networks help mitigate such socio-emotional challenges for novices. Consistent with design frameworks on how to engage novices in contributing to peer productions  \cite{kraut2012building}, our findings show that bond-based connection can also foster novices' confidence in presenting themselves for feedback. 

Step D in Figure \ref{fig:feedback-relationship-journey} marks the stage where the writer finds small, close communities, where they often solicit feedback for in-progress work. Interestingly, these groups are usually not established for the purpose of feedback \textit{per se}, but for socializing writers with similar interests, opinions and, many times, marginalized identities. Prior research has drawn attention to such private fandom spaces, where users feel more comfortable with disclosing a marginalized identity \cite{dym2020social} as well as gather together to overcome discrimination and hate \cite{fiesler2017growing,fiesler2016archive,fiesler2020moving}. Our findings suggest that the welcoming and safe environment that supports vulnerable identities may also help alleviate the vulnerability associated with presenting early-stage work for feedback. Writers built trust through a reciprocal process of socialization over time and personal disclosure (Step E in Figure \ref{fig:feedback-relationship-journey}). They turned to their trusted connections for in-progress feedback, support when they were stuck, and critique on high-fidelity work, because this trust helped them overcome emotional vulnerability in the creative process. While previous research has explored the possibility to generate scalable, specifically critical feedback from anonymized feedback providers \cite{hui2015using}, our study raises the question of whether such critiques can meet the emotional needs of creators.

\subsection{Feedback and the ecology of online spaces}
\label{discussion:ecology}
Our study finds an interrelationship between feedback and online space. Individual relationships between writers and feedback providers can move from public to private online spaces, and multiple social channels are involved in the creative process, often simultaneously. This is different from the migration of fandoms from one platform to the next, a long lasting phenomenon in fanfiction communities identified by \citet{fiesler2020moving}, which occurs due to evolving opinions, norms, and discrimination toward certain groups at a community level. We also observed platform migration as public, structural hostility towards queer creators persisted during our interviews in the form of the Tumblr NSFW ban. Our participants discussed how bans and purging of queer content had uprooted their online communities:

\begin{quote}

``When Tumblr banned not safe for work, it was really distressing for a bunch of us who don’t really fit on the very heteronormative sexual scale. So there was a lot of trying to figure out where we were going to go now, where we were, how would we stay connected, how would we continue to figure out and find stuff that we enjoyed.'' (P25)

\end{quote}

\noindent As a result, many writers moved to small, private online spaces to continue sharing and exchanging feedback on their creation. 

While platform migration is disruptive to feedback exchange, the everyday use of multiple platforms together can be beneficial, our finding suggests that feedback exchange, and the social activities that nourish effective feedback exchange, happen across multiple public and private platforms, echoing social computing literature that indicates that individuals tend to use an ecology of social media platforms at the same time \cite{zhang2017understanding,gonzales2015towards}. While prior research has begun to explore how different types of feedback can be solicited from different platforms \cite{yen2016social}, we elaborate that creators intentionally choose platforms and adjust their behaviors accordingly to seek feedback and connect with feedback providers.

Specifically, we learned from our participants that large, public fanfiction archives and event servers afforded writers with one-to-many social messaging such as reviews, access to fics written by others who share interests, opportunities to meet new people, and potential for discovering feedback providers. Open social media platforms such as Tumblr were used for advertising and reaching out to different populations. Private messaging channels like Discord, on the other hand, provided writers with smaller and more intimate spaces for discussing half-baked ideas and developing authentic relationships with feedback providers. Collaborative writing services like Google Docs supported writers in intensive editing and back-and-forth discussion with beta readers.

In particular, our findings point to the importance of close-knit communities and private spaces in online feedback exchange. 
Writers gravitated towards small community spaces, where they found the safety both to express themselves and receive critique in feedback that they could be sure was well-intentioned and kind. This was especially important for writers taking on topics that were marginalized in mainstream media. In other words, we observed that small, close communities are the spaces where transformative work happens. This need for closed groups in online fandom, especially for those who create media that challenges cisheteronormative narratives, traces back to pre-internet and early-internet ethnographic research. Hostility towards women in general and homoerotic works specifically, both before the internet and in its early days, spurred the creation of small, private counter-spaces where women would create slash fiction \cite{bacon1992enterprising, bury2005cyberspaces}.

Other than providing necessary safety conditions for challenging norms surrounding gender and sexuality, we found that private group chats were a crucial part of creative support for fanfiction writers. The safety they felt promoted relationship-building by making it possible for people to disclose more about their interests, identities and personal stories. This created common ground that lowered social barriers, motivated feedback providers, and established trust while giving and receiving critique, generating ideas, and seeking encouragement and support.

While many studies on online feedback focus primarily on large, open online communities \cite{cheng_critique_2020, kou2017supporting, xu2012what}, we argue that future researchers should pay more scholarly attentions to small close spaces, such as group chats, email threads, shared documents, and so on. Researchers could investigate these niche spaces to understand marginalized creators and their feedback and creative practices that are otherwise invisible to the public and support their needs. Additionally, existing studies of online feedback largely focus on a single platform, which dismisses the fact that feedback exchange naturally happen across an ecology of social channels. As our findings shed light on creators' feedback practices and choices across a range of different platforms, we hope to call for future research and design to engage diverse technical and social affordances, as well as the distinct types of social relationships fostered by different platforms in the feedback exchange process.

\subsection{Design Considerations: supporting affinity and connection-building in online feedback systems}
\label{discussion:implications}
%while no existing crowd-based or peer-based feedback systems can support the full feedback exchange process end-to-end. 
%Designers should consider affordances for connection and relationship-building in online feedback exchange systems. A repeated theme we found in relationship-building was moving from public asynchronous commentary to private synchronous chat, as well as exchanging social media handles. A feedback exchange system that facilitates this social media “handshake” could better fit into the ecosystem and support creators. This could be as simple as making the reviewer's social media handle salient on the page containing the review.

%The idea of initiating a cold reach-out over synchronous private chat was initially anxiety-inducing for many of the writers we spoke with. Those who were able to overcome this barrier reported that their anxiety was unfounded and that other writers were overwhelmingly friendly and encouraging. One potential design to help lower the social anxiety barrier could be allowing writers to indicate whether they would welcome a cold reach-out. However, lowering this barrier remains an open question that future research should explore.

Drawing from our insights presented above, we delineate the design opportunities for designers and researchers of online feedback systems to consider, especially regarding socio-emotional factors in feedback exchange. Albeit the limitation that not all findings from fanfiction communities may extend to every creative community, we urge designers and researchers to consider the following implications of our study as areas for future exploration in their respective feedback-exchange communities and platforms.

\subsubsection{Address a range of social needs in feedback}
Our study identified different social needs in feedback at different stages of the creative process. This finding indicates that designers of future systems should not build a one-size-fits-all solution for feedback exchange. Instead, systems should account for differing needs with flexible options in terms of selecting the right audience and communication channel. For example, users should be able to adjust the number of feedback providers that they want, choose between communication channels that are private or visible in public, and decide whether they want feedback from online strangers or people they know. Another possible design direction is that systems should be able to guide users towards strategies that meet their feedback needs, perhaps based on their activity histories and which stage they are at. For instance, the system could suggest feedback providers that have worked effectively with the user in the past when they need feedback on high-fidelity drafts, or recommend that the user broadcast their creation to a broader audience when they finish the work.

\subsubsection{Help feedback seekers signal interests and identity.}
Affinity in interests and identities motivated fanfiction writers to effectively exchange feedback. Future systems should support users to express their interests and identity in their feedback requests and connect them with compatible feedback providers. In particular, lessons can be taken from the design of Archive of Our Own, which prompts authors to attach signals of their interest and identity to their creations, in the form of free form authors' notes and checklist of user-created tags, so that their work could be seen by people who are interested in similar topics and hold similar views \cite{fiesler2016archive}. Fanfiction writers carefully select tags in order to make their work visible in public to the right audience of potential feedback providers. Future systems could consider similar approaches and integrate mechanisms such as reflective practices \cite{schon1984reflective} and visual categories \cite{xu2021ideaterelated} to help creators communicate interests and identity. 
%
%A common practice to signal affinities among fanfiction authors is to add appropriate tags to their fictions so that their work could be seen by people who have the similar interests or opinions. 
%Designers of feedback systems should consider how they might support creators in finding and communicating their affinities with audiences. 

\subsubsection{Support authentic relationships in feedback exchanges.}
Fanfiction writers preferred feedback providers with whom they had persistent, authentic and supportive relationships, as they knew the writers' previous work and goals, could balance emotional support with critical feedback, and had consistently provided quality feedback along a known domain of expertise. Designers of future systems should take this need into account and support formation and growth of authentic relationships between feedback seekers and providers. Fanfiction writers cultivated practices to identify and form connection for feedback, such as commenting on every work they read, replying to every comment they received, and normalizing cold reach-outs over direct message. Designers of feedback systems should consider how they might promote similar norms and afford access for people to reach out and connect in both public and and private channels. 

Further, fanfiction writers were able to deepen these relationships online through mutual social interaction, self-disclosure, and feedback exchange over time and across multiple communication channels. Future feedback systems should strive to scaffold these activities to support authentic relationships.  To do so, systems could encourage repeated feedback interactions, perhaps through offering the option for users to work with feedback providers repeatedly, supporting long-term relationship development. Systems could also create channels for synchronous socialization or off-topic discussions alongside feedback exchange so that users can feel they are interacting with ``real people'' \cite{ford2018we}. In addition, instead of forcing users to stay anonymous, systems could offer options for users to use consistent pseudonyms and express their personality, helping users recognize each other and feel authentic.

\subsubsection{Build inclusive, safe spaces for feedback.}
All of the above require a comfortable and safe environment. Future feedback systems should provide creators with a comfortable space so that they can safely be vulnerable, such as sharing early-stage work and reacting to criticism. The private help room presented in \citet{ford2018we} is a great example of such design. Furthermore, social computing designers and researchers should consider historical exclusivity towards marginalized identities in the design of online spaces, and build spaces that include creators of all identities. Modeling after communities of fanfiction writers, designers may consider the role of semi-public and private spaces in creating safety. Crucially, access to these spaces must be controlled by community members. In addition, designers of public feedback spaces may embrace inclusivity as a design value and instantiate it in the design of the platform \cite{fiesler2016archive}.

\subsection{Implications for Fanfiction Writers}
We are well-aware that there is an overlap between social computing researchers, fanfiction writers, and fanfiction community practitioners. We hope the insights from our study could pay back to the fanfiction writing community, especially for those who seek feedback on their work and those who offer feedback to others. Writers looking for making connections with feedback providers may consider the strategies described in this paper, such as attending community events, starting discussions in public comments, and seeking out small online spaces. Writers may also ``just-reach-out'' to strangers in the same fandom or interest group for feedback, as most of our participants agree that affinity makes the community welcoming and considerate. In addition, practitioners who are positioned more centrally in the community may consider how they can help other members, especially newcomers, identify feedback resources and establish feedback relationships in the community. Strategies to consider can include warmly responding to reach-outs, organizing events, adopting norms such as reciprocity in comments, and cultivating safe spaces for writers to engage in transformative work. The many powerful stories we heard in our interviews are evidences that these practices make a difference.

\section{Limitation \& Future Works}
While we believe that our study contributes several insights about feedback exchange in online affinity networks, we admit that a number of aspects of our study can be improved or extended. 
First, we chose interview as our method for investigating feedback practices (and more importantly, the motivation and rationales behind those practices). Due to the nature of our recruitment strategy, all of our participants were self-selected and unpaid, which to some extent implied that they were intrinsically motivated to speak about their experiences with feedback exchange in the community. Therefore, we did not have a chance to gain insight from writers who had overall negative experiences in the fanfiction communities, dropped out from fanfiction altogether, or were never able to engage in an effective feedback experience. This may contribute to a bias in our findings towards positive aspects of feedback exchange. 

Another limitation derived from our recruitment strategy is that our study demographics disproportionately consist of certain groups. Likely due to the fact that our recruitment survey were written in English, all of our participants were from the United States or European countries, with the majority of them from white-dominant English-speaking countries. Due to this reason, 21 out of the 25 participants who reported their race were white. Therefore, it is possible that our insights were biased towards this population and that we were not able to identify experience and needs specific to a different group. We want to point out that we did put in our best effort and were able to recruit participants with a diversity of genders\textemdash 14 participants were identified as women and the rest as other nonconforming genders. This matches the population of online fanwork communities and serves as a response to the lack of gender diversity and the representation of the queer communities in general HCI studies \cite{linxen2021weird}. Nevertheless, we were not able to recruit any cis men participants. Future research can extend this work and test our findings in a broader population of fanfiction writers. 

Our work is further constrained in that we recruited participants from a specific platform\textemdash Tumblr. Although Tumblr was one of the largest platforms for fanfiction writers to share their work and socialize at the time of our study (January to February, 2019), due to their changes in content policy in late 2018 \footnote{\url{https://www.businessinsider.com/tumblr-bans-nfsw-content-and-users-say-the-platform-will-suffer-2018-12}}, there has since been a significant community migration away from the platform. There is a possibility that feedback-related behaviors associated with Tumblr may have been changed since we conducted our study. Luckily, most of our participants used more than one platform or channel to present their work, seek feedback, and connect with feedback providers, so many of our findings should still hold even though Tumblr has largely changed. 

Finally, we note that the present study is a case study with online fanfiction writers. While online fanfiction communities have been a prototypical example for the study of online affinity, we cannot guarantee that our findings about feedback-seeking among fanfiction writers are generalizable to all types of creators in all types of affinity networks. Future work should explore whether similar practices exist among other similar or different creators in online affinity spaces or networks. In addition, although we were able to identify prominent feedback practices, we did not study the prevalence of these themes due to the nature of interview and our choice of grounded theory as our analysis method. Future research can build on our work and quantitatively measure the prevalence of these practices. 

\section{Conclusion}
Fanfiction writers were some of the earliest adopters of the internet \cite{coppa2006brief}. This ever-evolving community of communities has transformed media consumption and production \cite{jenkins2006convergence}, inspired new understandings about how shared interests drive learning \cite{ito2018affinity}, and innovated in the design of internet spaces to incorporate values and norms around accessibility, inclusivity and identity \cite{fiesler2016archive}. Social computing researchers interested in these topics can learn quite a bit from fanfiction communities, and indeed, conducting this set of interviews has changed the authors' approach to their research. The stories we heard from fanfiction writers helped us believe that the internet can be designed to be better than it is now, a place for creators of all types of media from all backgrounds to learn from each other. Our grounded theory of feedback-seeking in fanfiction communities expanded knowledge about the role of feedback throughout the creative process and established an interrelationship between online connection and feedback. We discussed strategies writers used and provided future directions for research and design around online feedback exchange systems. We recommend that designers and researchers of online creative spaces consider how they will afford connections between people of like interests, support the growth of authentic connections and instill values of reciprocity and inclusivity. We also encourage readers who are passionate about the topic of designing inclusive, authentically connected communities for learning to ``just reach out'' to us and continue the conversation.

\section{Acknowledgments}
Thank you to all of the fanfiction writers who enthusiastically participated in this all-volunteer study. You trusted us with raw, powerful experiences of connection and community, and transformed the way we think about feedback. Thank you to Ruby Davis, who supported us throughout this research and greatly contributed to the sourcing effort for this study. Thank you to Cecilia Aragon for your support over the years we worked on this.

\bibliographystyle{ACM-Reference-Format}
\bibliography{sample-base}

\appendix

\end{document}